\documentstyle[aps,prl,epsf,multicol]{revtex}
\newcommand{\be}{\begin{equation}}
\newcommand{\ee}{\end{equation}}
\newcommand{\ba}{\begin{array}}
\newcommand{\ea}{\end{array}}
\newcommand{\bea}{\begin{eqnarray}}

\newcommand{\eea}{\end{eqnarray}}

\newcommand{\ra}{\rightarrow}

\begin{document}
\draft
\title{Fluctuations of self-flattening surfaces}
\author{Yup Kim$^{1,2}$, S. Y. Yoon$^1$ and Hyunggyu Park$^{3,4}$}
        
\address{$^1$ Department of Physics and Research Institute of Basic Sciences, 
         Kyung Hee University, Seoul, 130-701, Korea}
\address{$^2$ Asia Pacific Center for Theoretical Physics, Korea}
\address{$^3$ Department of Physics, Inha University,
         Inchon 402-751, Korea}
\address{$^4$ School of Physics, Korea Institute for Advanced Study,
         Seoul 130-012, Korea}
\maketitle

\begin{abstract}
We study the scaling properties of self-flattening surfaces
under global suppression on surface fluctuations. 
Evolution of self-flattening surfaces is described by
restricted solid-on-solid type monomer deposition-evaporation model 
with reduced deposition (evaporation) at the globally highest (lowest) site.
We find numerically that equilibrium surface fluctuations are anomalous 
with roughness exponent $\alpha\simeq 1/3$ and dynamic exponent 
$z_W \simeq 3/2$ in one dimension (1D) and $\alpha=0~ (\log )$ and 
$z_W \simeq 5/2$ in 2D.  Stationary roughness can be understood 
analytically by relating our model to the static self-attracting random walk model 
and the dissociative dimer type deposition-evaporation model.
In case of nonequilibrium growing/eroding surfaces, 
self-flattening dynamics turns out to be irrelevant and the normal 
Kardar-Parisi-Zhang universality is recovered in all dimensions.
\end{abstract}


\begin{multicols}{2}

Structural properties for fluctuating surfaces under thermal noise
have been studied extensively\cite{dyn}. Equilibrium surfaces with proper 
surface tension are always rough in one dimension (1D) and display 
a roughening transition in two dimensions (2D)\cite{weeks}. Higher dimensional
surfaces are always smooth.
Surface roughness is well documented and classified as
the Edwards-Wilkinson (EW) universality class\cite{EW}.
The EW class is generic and robust for equilibrium surfaces with local 
surface tension. Only specific nonlinear contributions in nonequilibrium 
growth processes may become relevant and drive the system into
other universality classes, e.g., the Kardar-Parisi-Zhang (KPZ)
universality class\cite{KPZ}
.

In this paper, we introduce a new global mechanism to suppress 
surface fluctuations, besides ordinary local surface tension. 
We call it {\em self-flattening} mechanism to reduce
growing (eroding) probability at the globally highest (lowest) point 
on the surface. This global type suppression makes the surface less rough,
which may bring forth new universality classes for equilibrium and
nonequilibirum surfaces. Inclusion of suppression at all local
extremal points leads to less interesting layer-by-layer growth processes 
and the steady state surfaces are always smooth with finite fluctuation width.


We describe surface configurations in terms of integer height variables 
$\{ h ({\vec r}) \}$ at site ${\vec r}$ on a $D$-dimensional 
hypercubic lattice. They are subject to the restricted solid-on-solid 
(RSOS) constraint, $h(\vec r +{\hat {e_i}}) -h(\vec r)= 0, \pm 1$ 
with $\hat {e_i}$ a primitive lattice vector in the $i$-th direction ($i=1,...,D$).
The RSOS constraint effectively generates local surface tension which
prevents indefinite growth of surface fluctuations for finite systems.

Evolution rule for the ordinary RSOS type monomer 
deposition-evaporation model is given as follows. 
First, select a site ${\vec r}$ randomly. Next, deposit a particle,
$h({\vec r}) \rightarrow h({\vec r}) + 1$, with probability $p$ or
evaporate a particle, $h({\vec r}) \rightarrow h({\vec r}) - 1$, with
probability $q=1-p$. Any deposition/evaporation attempt is rejected 
if it would result in violating the RSOS constraint. Equilibrium surfaces at $p=q$ belongs
to the EW class, while nonequilibrium growing/eroding surfaces at $p\neq q$
the KPZ class\cite{dyn,KK}. 

For self-flattening surfaces, we need a slight variation of the
evolution rule to incorporate the global suppression :
only when deposition (evaporation) is attempted at the globally highest (lowest) site,
the attempt is accepted with probability $u$ and rejected with probability $1-u$.
At $u=1$, the ordinary RSOS model is recovered. 
The $u=0$ case is trivial, because the surface is confined within initial surface width. 

We perform numerical simulations, starting from a flat surface 
of linear size $L$ with periodic boundary conditions. We measure the surface fluctuation
width $W$ as
\begin{equation} 
W^2 (L,t)={1\over L^D} \sum_{{\vec r}} \left\langle \left[ h({\vec r},t)- 
{1\over L^D}\sum_{{\vec r}} h({\vec r},t) \right]^2 \right\rangle ,
\label{W2} 
\end{equation}
where  $\left\langle \cdots \right\rangle$
represents the ensemble average with equal weights. Therefore,
our simulations at $p=q$ correspond to the infinite temperature limit 
of equilibrium RSOS surfaces. 
The surface width satisfies the dynamic scaling relation 
\begin{equation} 
W(L,t)=L^\alpha f\left({t / L^{z_W}}\right), \label{scaling} 
\end{equation}
where the scaling function $f(x)\ra {\rm const.}$ for $x\gg 1$ and 
$f(x) \sim x^\beta$ $(\beta =\alpha/z_W)$ for $x\ll 1$\cite{dyn,FAV}. 

First, we report the numerical results for equilibrium surfaces ($p=q$).
For 1D, we run simulations for $L=2^5, ..., 2^{11}$ at 
$u=0.1$, 0.3, 0.6, and 0.8, and average over at least 300 independent samples.
In early time regime ($t\ll L^{z_W}$), the surface width grows with time, 
$W\sim t^\beta$, and saturates to a finite value which increases with size, 
$W_s \sim L^\alpha$. 

In Fig.~1, we show the plot of  $\ln W$ versus $\ln t$ at $u=0.6$ 
for various system sizes. The growth exponent $\beta$ is estimated by a simple 
straight line fitting of early time data for the largest system size $L=2^{11}$. 
Our estimate is $\beta=0.22(1)\simeq 2/9$. We also analyze the data
at other values of $u$ and find that $\beta$ does not vary with $u$.

In order to extract the stationary property,
we average over data in the saturated regime ($t\gg L^{z_W}$) for given $L$ 
to estimate $W_s (L)$.  For efficient estimation of $\alpha$, we introduce effective 
exponents
\begin{equation}
\alpha_{eff} (L)=\ln \left[ W_s (mL)/W_s(L) \right] / \ln m, \label{effective}
\end{equation}
where $m$ is an arbitrary constant (here, we set $m=2$). 

Effective exponents at various values of $u$ 
are plotted in Fig.~2. Close to $u=1$, our data show large corrections to
scaling as expected, due to the presence of the EW fixed point 
($\alpha=1/2, \beta=1/4$) at $u=1$. However, the asymptotic estimates
seem to be independent of $u$. We estimate that
$\alpha=0.33(1)\simeq 1/3$ for all $u$.
We check the dynamic scaling relation directly by plotting $W/L^{\alpha}$
versus $t/L^{z_W}$ in the inset of Fig.~1. 
Our data collapse very well with $\alpha=1/3$
and $z_W=3/2$ for all $u$, which are consistent with the above results.

This set of scaling exponents form a new universality class, distinct from
the EW and any previously known growth-type universality class. 
It implies that the self-flattening dynamics is a relevant perturbation
to the EW fixed point in 1D. Therefore, the continuum equation to describe 
self-flattening surfaces must contain a global-type nonlinear term. 
Further study in this direction is left for future research.

In case of 2D EW surfaces, it is well known that the surface width grows
logarithmically with time and its saturated value also increases logarithmically
with size\cite{weeks}. Especially, the saturated width $W_s$ scales for large $L$ as 
\begin{equation}
W_s^2 (L)\simeq {1 \over {2\pi K_G }} \ln L, \label{KGs}
\end{equation}
where $K_G$ is the effective coupling constant 
of the Gaussian model where equilibrium surface models
flow into by renormalization group transformations\cite{weeks,den}. 
The ordinary RSOS model at the infinite temperature (our model at $u=0$)
is known to take $K_G=K_G^0\simeq 0.916$\cite{den,LD}.
 
Assume the dynamic scaling relation similar to Eq.(\ref{scaling}) as
\begin{equation}
W^2 (L,t)= {1 \over {2\pi K_G }} \ln \left[L~ g\left({t / L^{z_W}}\right) \right],
\label{scaling2}
\end{equation}
where the scaling function $g(x)\ra {\rm const.}$ for $x\gg 1$ and 
$g(x) \sim x^{1/z_W}$ for $x\ll 1$.  Then, in early time regime ($t\ll L^{z_W}$), 
the surface width grows as
\begin{equation}
W^2 (t)\simeq {1 \over {2\pi K_G z_W}} \ln t . \label{KGt}
\end{equation}
The amplitude ratio in Eqs.(\ref{KGs}) and (\ref{KGt}) yields 
a value of the dynamic exponent $z_W$.
The EW surfaces take $z_W = 2$ in all dimensions.


We run simulations on $L\times L$ lattices with $L=2^3, ..., 2^7$
at $u=0.1$ and 0.5 and average over at least 300 independent samples.
In Fig.~3(a), we plot $W^2$ against $\ln t$ at $u=0.5$. It shows a nice linear behavior
in the early time regime.
In Fig.~3(b), we plot $W_s^2 $ against $\ln L$, which also shows
a very nice linear behavior. We measure its slope and find that 
$K_G \simeq 0.92(1)$ for all $u$, which is very close to $K_G^0$. 
In contrast to the 1D surfaces, the global suppression does not seem to change
the asymptotic behavior of the stationary surface roughness. As can be seen in Fig.~3,
it seems to shift $W_s$ only by a constant.

We measure the amplitude ratio by comparing two slopes in Figs.~3(a) and (b).
We estimate $z_W = 2.5(1) \simeq 5/2$ for all $u$, which is clearly
distinct from the EW value of 2. We also check the dynamic scaling relation of 
Eq.(\ref{scaling2}) by plotting $W^2-W_s^2$ versus $t/L^{z_W}$ in Fig.~3(a).
Our data collapse reasonably well with $z_W=5/2$ for all $u$.  
Together with our 1D results, we conclude that the self-flattening surfaces 
display a new type of scaling behavior and form a novel universality class.

The partition function for equilibrium self-flattening surfaces can be written as 
\begin{equation}
Z=\sum_{\rm RSOS ~conf.} e^{-\beta\left(h_{\rm max} -h_{\rm min}\right)} ,
\label{partition}
\end{equation}
where the summation is over all height configurations satisfying 
the RSOS condition, $\beta$ a temperaturelike parameter, 
and $h_{\rm max}$ ($h_{\rm min}$)  the
globally maximum (minimum) height for a given configuration.

Global suppression for self-flattening dynamics is simply Metropolis
type evolution algorithm with this partition function to reach the equilibrium. 
Deposition (erosion)  at the globally highest (lowest) site increases 
the energylike term $h_{\rm max} -h_{\rm min}$ by one unit and 
these attempts are accepted with Boltzmann type probability $e^{-\beta}$.
Any other deposition (erosion) attempts are always accepted,
because they do not increase the energylike term. 
Of course, all attempts resulting in violation of the RSOS constraint are
rejected. By identifying $u=e^{-\beta}$, our model for self-flattening surfaces is
exactly the same as Metropolis evolution with the above partition function.

Stationary property of this system can be understood analytically.
In 1D, this system is equivalent to the so-called {\em static self-attracting
(timid) random walks} \cite{SARW}. The surface can be mapped to the time trajectory
of a random walker by identifying the height $h(x)$ at site $x$ with
the walker position after $x$ steps. The system size $L$ becomes the
total number of steps and the RSOS constraint limits one-step hopping 
distance to 0 or $\pm 1$. 

In 1D, the energylike term is simply the number of distinct sites visited
by the random walker up to $L$ steps. Random walk configurations 
with less visited sites are preferred. Such a random walker
tends to visit previously visited sites, so
the walk is self-attractive. Its typical displacements 
are known rigorously to scale as $L^{1/(D+2)}$ \cite{SARW,DV} 
under the assumption that the visited sites form a compact cluster. 
In 1D, the cluster is obviously compact,
so the roughness exponent in our model should be $\alpha=1/3$ in 1D.

In 2D, the self-flattening surfaces are completely different from the self-attracting 
walks. The former deals with membrane fluctuations, while the latter polymer
fluctuations. In order to understand the scaling behavior of the self-flattening
surfaces, we investigate the intricate relation between our model and the
dissociative dimer deposition-evaporation model in equilibrium \cite{LD,NKPD}.

In the dimer model, we deposit or evaporate particles only in a dimer form
aligned along the surface. There is a global {\em evenness} conservation law 
that the number of
particles at each height level must be conserved modulo 2\cite{NKPD}. 
This leads to 
a Boltzmann type factor in the partition function as
\begin{equation}
Z=\sum_{\rm RSOS ~conf.} \prod_h\frac{1}{2}\left(1+z^{v_h}\right),
\label{partition2}
\end{equation}
where the product is over all possible height levels and $v_h$ 
the number of particles at height level $h$. The dimer model 
corresponds to the $z=-1$ case where only configurations
obeying the evenness conservation law (all $v_h$ are even)
survive in the partition function.
At $z=1$, the model reduces to the ordinary monomer model.

The self-flattening surfaces correspond to the $z=0$ limit.
Each term inside the product picks up a factor of $\frac{1}{2}$ if $v_h\neq 0$,
otherwise a factor of unity. The number of height levels with nonzero $v_h$
(at least one particle) is $h_{\rm max} -h_{\rm min}$. Therefore, the
$z=0$ case is equivalent to the self-flattening surfaces at $\beta=\ln 2$.
In fact, the $Q$-mer generalization corresponds 
to the $\beta=\ln Q$ case \cite{NKPD,KKP}.

>From the Gaussian model type renormalization group argument,
one can show that  the 2D surface roughness is always logarithmic in the
dimer model for $-1\le z<1$ (see Eq.(\ref{KGs})) and its amplitude 
remains unchanged \cite{LD}. Our numerical results for all $u$ are
consistent with this. The dimer characteristics show up only 
in a form of corrections to scaling. Recently, it is suggested that the corrections 
to scaling should scale as $\ln (\ln L)$, which is confirmed for the dimer 
model at $z=-1$\cite{LD}. We find no evidence of this type of corrections to 
scaling in our model ($z=0$) and the leading corrections are constants. 
The origin of this discrepancy between the $z=0$ and $z=-1$ case
is not fully understood as yet.

Next, we consider nonequilibrium growing/eroding surfaces $(p\neq q)$.
We run simulations for $L=2^5, ..., 2^{11}$ for 1D and $L=2^3, ..., 2^7$ for 2D
at $p=1$ with $u=0.5$ and $u=1$ (ordinary RSOS). In Fig.4, we plot $\ln W_s$ 
against $\ln L$ and,
in the inset, $\ln W$ against $\ln t$ for the largest system size in 1D and 2D, respectively. 
We do not find any noticeable change of $W$ ascribed to the global suppression.
We estimate that $\alpha\simeq 0.50(1)$ and $\beta\simeq 0.32(1)$ for 1D and
$\alpha\simeq 0.40(1)$ and $\beta\simeq 0.24(1)$ for 2D, which are consistent
with the results for the ordinary RSOS model \cite{KK}. 
We conclude that the global suppression is irrelevant to
nonequilibrium growing/eroding surfaces.

In summary, we studied the scaling properties of the self-flattening surfaces
in 1D and 2D. Equilibrium surfaces display dynamic scaling behavior distinct
from the EW class and form a new universality class. We show that stationary roughness
can be understood through mapping our model to self-attracting random walks in 1D and 
dissociative dimer type deposition-evaporation model in 2D. In higher dimensions,
the surfaces are always smooth. In contrast, nonequilibrium self-flattening surfaces 
belong to the ordinary KPZ universality class. This implies that
the self-flattening dynamics is strong enough 
to dominate over the EW type local surface tension term, but weaker than the KPZ type nonlinear term.
It would be very interesting to find a continuum-type equation to govern the self-flattening dynamics. 

We thank Marcel den Nijs, Deok-Sun Lee, and Lucian Anton for useful discussions. This research is supported 
in part by  by grant No.~R01-2001-00025 (YK) and by grant No.~2000-2-11200-002-3 [HP]
from the Basic Research Program of KOSEF.

\begin{figure}
\centerline{\epsfxsize=70 mm  \epsfbox{Fig1.epsi}}
\vskip 10 true pt
\caption{Plots of $\ln W$ against $\ln t$ for 1D self-flattening 
equilibrium surfaces at $u=0.6$. The slope of the straight line 
is $\beta=0.22(1)$. 
The inset shows the data collapse with $\alpha=1/3$ and $z_{W}=1.5$.}
\end{figure} 
 
\begin{figure} 
\centerline{\epsfxsize=70 mm  \epsfbox{Fig2.epsi}}
\vskip 10 true pt
\caption{Effective exponents $\alpha_{eff}$  versus $1/L$ for 1D self-flattening 
equilibrium surfaces. All data for various values of $u$ converge to $1/3$ rather nicely 
in the $L \rightarrow \infty$ limit.}
\end{figure} 

\end{multicols}

\begin{figure}
\centerline{\epsfxsize=70mm {\epsfbox{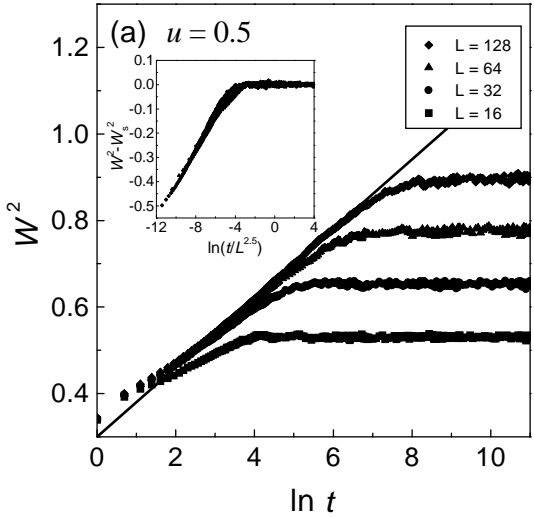}}\hspace{1cm}\epsfxsize=70mm {\epsfbox{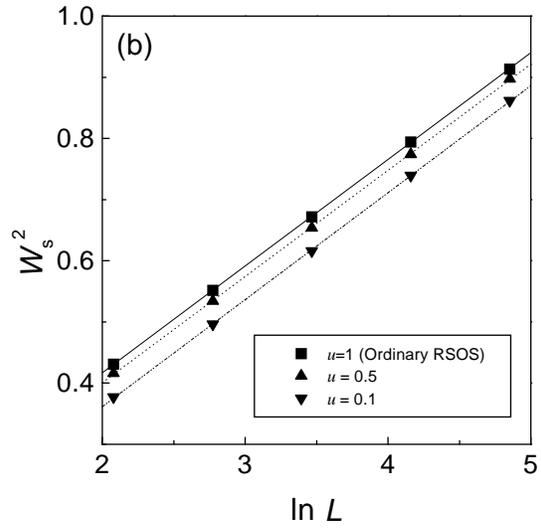}}}
\vskip 10 true pt
\caption{(a) Plots of $W^2$ against $\ln t$ at 
$u=0.5$ for 2D self-flattening equilibrium surfaces. 
The slope of the straight line yields the value of 
$K_G z_W = 2.3(1)$. The inset shows the data collapse
with $K_G=0.92$ and $z_W=5/2$.
(b) Plots of ${W_s}^2$ against $\ln L$ at $u=1$
(ordinary RSOS), $u=0.5$, and $u=0.1$.
The slopes of three straight lines yield the same value of 
$K_G=0.92(1)\simeq K_G^0$.}
\end{figure} 

\newpage
\begin{multicols}{2}

\begin{figure}
\centerline{\epsfxsize=70mm \epsfbox{Fig4.epsi}}
\vskip 10 true pt
\caption{Plots of $\ln {W_s}$ against $\ln L$ in 1D and 2D
nonequilibrium growing surfaces at $p=1$. 
There is no noticeable difference in $W$ between the $u=1$ (ordinary RSOS) and 
$u=0.5$ (self-flattening) case. In the inset, we plot early time behavior of $W$ for 
system sizes $L=2^{11}$ (1D) and $L=2^7$ (2D). Straight line fits yield $\alpha=0.50(1)$,  $\beta=0.32(1)$ for 1D and 
$\alpha=0.40(1)$, $\beta=0.24(1)$ for 2D.}
\end{figure} 

\end{multicols}

\end{document}